\documentclass[lettersize,journal]{IEEEtran}
\usepackage{cite}
\usepackage{amsmath,amssymb,amsfonts}
\usepackage{graphicx}
\usepackage{hyperref}
\usepackage{textcomp}
\usepackage{amsmath,amsfonts}
\usepackage{romannum}%
\usepackage[linesnumbered,ruled,vlined]{algorithm2e}
\usepackage{algpseudocode}
\usepackage{array}
\usepackage{subfig}
\usepackage{textcomp}
\usepackage{xcolor}
\usepackage{stfloats}
\usepackage{url}
\usepackage{verbatim}
\usepackage{graphicx}
\usepackage{hyperref}
\usepackage{cite}
\usepackage{orcidlink}
\usepackage{balance}
\hypersetup{hidelinks}

\def\BibTeX{{\rm B\kern-.05em{\sc i\kern-.025em b}\kern-.08em
    T\kern-.1667em\lower.7ex\hbox{E}\kern-.125emX}}
\SetKwInOut{Parameter}{parameter}
\begin{document}

\title{Deep Learning Enabled Segmentation, Classification and Risk Assessment of Cervical Cancer}

\author{Abdul Samad Shaik \orcidlink{0009-0008-3356-9697}, Shashaank Mattur Aswatha \orcidlink{0000-0001-5450-3568}, and Rahul Jashvantbhai Pandya \orcidlink{0000-0002-3259-817X}
\thanks{This work was supported in part by the Department of Telecom-
communication (DoT), Ministry of Communications, Government
of India under the Telecom Technology Development Fund
(TTDF) the scheme implemented through TCOE India under the
grant TTDF/6G/48, Science and Engineering Research Board (SERB) Project EEQ/2020/000047 and Project
SIR/2022/00095.}
\thanks{Abdul Samad Shaik, Shashaank Mattur Aswatha and Rahul Jashvantbhai Pandya are with the Department of Electrical, Electronics, and Communications Engineering, Indian Institute of Technology, Dharwad, India (e-mail: ee23mt018@iitdh.ac.in; matturas@iitdh.ac.in; rpandya@iitdh.ac.in).}}

\maketitle
\thispagestyle{empty}
\pagenumbering{arabic}
\begin{abstract}
Cervical cancer, the fourth leading cause of cancer in women globally, requires early detection through Pap smear tests to identify precancerous changes and prevent disease progression. In this study, we performed a focused analysis by segmenting the cellular boundaries and drawing bounding boxes to isolate the cancer cells. A novel Deep Learning (DL) architecture, the ``Multi-Resolution Fusion Deep Convolutional Network", was proposed to effectively handle images with varying resolutions and aspect ratios, with its efficacy showcased using the SIPaKMeD dataset. The performance of this DL model was observed to be similar to the state-of-the-art models, with accuracy variations of a mere 2\% to 3\%, achieved using just 1.7 million learnable parameters, which is approximately 85 times less than the VGG-19 model. Furthermore, we introduced a multi-task learning technique that simultaneously performs segmentation and classification tasks and begets an Intersection over Union score of 0.83 and a classification accuracy of 90\%. The final stage of the workflow employs a probabilistic approach for risk assessment, extracting feature vectors to predict the likelihood of normal cells progressing to malignant states, which can be utilized for the prognosis of cervical cancer.
\end{abstract}


\vspace{-5mm}
\section{Introduction}
\IEEEPARstart{C}{ervical Cancer}, characterized by the uncontrolled proliferation of cells in the cervix, rapidly emerges as a serious threat in oncology. This cancer is primarily seen among women over the age of 30 years and is on the rise worldwide. The persistent strains of human papillomavirus, a virus typically contracted during sexual intercourse, are the primary cause of this malignancy. So, early detection of cervical cancer is essential for its management. Regular screening is the cornerstone of early detection, and it has been proven to be essential to prevent and treat cervical cancer. Cervical cancer is detected early through Pap smear testing \cite{paptest}. In this test, samples collected from the cervix are examined by expert pathologists for any abnormalities. The analysis of abnormalities through manual testing relies on pathologists, but it is often time-intensive and prone to errors. Automation is essential to address these challenges. Artificial intelligence (AI) techniques have been extensively employed for automated cervical cell classification in recent years. Most of these existing AI techniques rely on feature extraction, often leveraging Machine Learning (ML) algorithms such as Support Vector Machine (SVM), K-Nearest Neighbor (KNN), and Artificial Neural Networks (ANN) etc. Although ML techniques perform well on smaller datasets, medical datasets are typically large and complex. Accurate ML classification can be challenging due to cell shape and size variations, and different feature extraction methods yield varying results. As a result, the need for automated feature extraction has increased, and Deep Learning (DL) techniques effectively address this. DL techniques are usually applied in several areas of medical imaging. Although DL has significantly advanced the detection of cervical cancer cells, most architectures are large, with millions of learnable parameters, requiring substantial computational time. Currently, there is a lack of lightweight models to overcome these challenges. This work uses lightweight DL methods to facilitate better cervical cancer cell diagnosis outcomes.

\vspace{-4mm}
\subsection{Related Works}
In cervical cancer detection and diagnosis, many investigations have explored multiple methodologies, including established screening methods, such as Pap smears, and the pioneering utilization of ML and DL frameworks.

Numerous studies have explored various ML approaches to automate the classification of cervical cancer from images. Another study \cite{smote} implemented a classification model using Random Forest (RF) by exploiting known risk factors associated with cervical cancer, where class imbalance was addressed with the SMOTE (Synthetic Minority Oversampling Technique) and dimensionality reduction was performed utilizing Principal Component Analysis (PCA)  \cite{pca_paper}. Nevertheless, ML-based methodologies have revealed inherent constraints, including suboptimal accuracy in distinguishing specific cell types and a reliance on manual feature engineering. This process is labour-intensive and prone to subjective bias and increased risk of errors \cite{ml_disadv}.
Costa \textit{et al.} \cite{10.1371/journal.pone.0138945} has made an extensive study from 2006 to 2013 and emphasizes the need for proactive measures to enhance the effectiveness of cervical cancer control in Brazil. This highlights the urgency for strategic interventions and improvements within the country's cervical cancer screening program to address the identified challenges and ensure better disease control.

DL has garnered significant triumph across diverse applications, particularly in cancer research. From 2016 to 2023, \cite{review_paps} explores advancements in segmentation and classification of Pap smear images for cervical cancer detection using DL techniques. It highlights the effectiveness of methods like UNet \cite{unet-seg}, Mask Region based Convolutional Neural Networks (RCNNs), SVM, and Convolutional Neural Network (CNN), achieving over 90\% accuracy, with hybrid techniques. CNNs are the leading DL architecture for image analysis, extensively applied in the segmentation and classification \cite{cnn_survey}. R. Ahmed \textit{et al.}\cite{latest_paps} presented a multi-deep transfer learning model integrating feature extraction from advanced architectures like MobileNet \cite{mobilenetv3-large-1} and ResNet-50 \cite{resnet}, PCA-based feature reduction, and a novel smoothing cross-entropy loss function for cervical cancer detection, achieving accuracy more than 90\%. M. Kaur \textit{et al.} \cite{9954901} proposed MLNet: Metaheuristics based lightweight DL network, trained and validated it across three benchmark cervical cancer datasets, and demonstrated a superior performance compared to the existing models. On the Herlev \cite{jantzen2008} dataset, MLNet demonstrated enhancements in accuracy, F-measure, sensitivity, specificity, and precision, with improvements ranging from approximately 1.5\% to 1.7\%. Similarly, on the SIPaKMeD \cite{sipakmeddataset} dataset, MLNet outperformed existing models by approximately 1.9\% to 2.2\%. In contrast, the Mendeley Liquid based Cytology (LBC) \cite{mendleylbc} dataset achieved better results by approximately 1.4\% to 1.6\% across these evaluation metrics. These findings highlight the consistent and noteworthy advancements of MLNet over established models for cervical cancer detection. L. Zhang \textit{et al.} \cite{article123} introduced a novel approach using a CNN for cervical cell classification, moving away from prior methodologies that rely on manual segmentation and hand-crafted features. This method autonomously extracted deep features from cell image patches by leveraging coarse nucleus-centered patches as the network input. Transferring pre-trained model features to a ConvNet and aggregating predictions achieved superior performance on the Herlev Pap smear, and H\&E stained manual liquid-based cytology datasets compared to the previous methods. There is potential for improving automation-assisted primary cervical screening systems with this segmentation-free and extremely accurate categorization approach. M. Fang \textit{et al.} \cite{adeepneuralnetwork} introduced the neural architecture named DeepCell, comprising three distinct foundational modules aimed at capturing feature intricacies through a diverse array of kernels with varying dimensions. Subsequently, these modules were iteratively amalgamated to construct a comprehensive model for classifying the cervical cells. Their approach exhibited remarkable performance metrics, with accuracy, recall, precision and F-score of 95.63\%, 95.65\%, 95.68 \% and 95.64\%, respectively, on the dataset SIPaKMeD, surpassing all the other methodologies. Recent advancements have continued this trend, integrating newer techniques for improved performance. In December 2024, a study introduced a method combining RES\_DCGAN for data augmentation with ResNet50V2 integrated with self-attention mechanisms. This approach enhanced classification performance of Pap smear images, addressing data scarcity challenges and improving generalization capabilities \cite{res_dcgan_resnet50v2}. Similarly, Di Piazza and Boussel \textit{et al.} \cite{ps3c_framework}  proposed PS3C, an ensemble-based two-step framework. It first filtered out unsuitable images using a neural network and then classified the remainder as healthy, unhealthy, or both, aiming to assist cytologists in managing the increasing workload from Pap smear screenings. Additionally, a 2025 study \cite{transfer_learning_comparison} conducted a thorough evaluation of 16 pre-trained CNN models on the Herlev and SIPaKMeD datasets, revealing ResNet50’s effectiveness (95\% accuracy on both 2-class and 7-class Herlev tasks), while VGG16 achieved 99.95\% on SIPaKMeD .

Using DL methodologies, anomalous cells have been segmented from conventional Pap smear digital images \cite{deeplearning123},\cite{deeplearning1234}. Y. Song \textit{et al.} \cite{Song2014ADL} introduced a method for segmenting cervical cytoplasm and nuclei through the integration of superpixels and CNNs. Y. Liu \textit{et al.} \cite{8468154123} presented an approach utilizing Mask RCNNs for the automatic segmentation of cervical nuclei coupled with the local fully connected conditional random field. J. Huang \textit{et al.} \cite{95132829513282}  introduced the Cell-Generative Adversarial Network (GAN) methodology, which underwent evaluation in segmenting both single-cell and overlapping-cell images. They reported a 94.3\% Dice Coefficient (DC) and a 7.9\% False Negative Rate (FNR) overhead for single-cell images. In the case of overlapping cell images, they achieved 89.9\% DC and 6.4\% FNR values. Y. Song \textit{et al.} \cite{75624007562400} introduced another DL model to tackle segmentation challenges, particularly in images featuring numerous overlapped cells. Their method provides a more precise delineation of abnormal cells within densely overlapped images. However, the performance deteriorated when the overlapping affected the nucleus. The Levenberg–Marquardt feedforward Multi Layer Perceptron (MLP) neural network was employed \cite{articlearticle1} to categorize cervical images from a cohort of one hundred patients. In \cite{wumiao123}, various DL techniques were used to perform cervical cell classification. Numerous research studies concerning the segmentation and classification of the nucleus have been examined within this section. For a comprehensive classification of cervical cells, besides the nucleus data, the segmentation of the entire cell is seen to be more appropriate
\cite{chankong2014automatic,li2012cytoplasm,pai2012nucleus,tsai2008nucleus}. In segmentation-based approaches, the properties of the segmented regions are then classified using different classifiers.

In this article, we perform segmentation and classification tasks and Risk Assessment (RA) on a dataset, SIPaKMeD \cite{sipakmeddataset}, comprising Pap smear images. To the best of the authors' knowledge, this work represents the first attempt to address these challenges with a unique approach. A novel lightweight DL architecture has been designed specifically for classifying cervical cells. A multi-task learning technique has been proposed to simultaneously perform segmentation and classification tasks on cervical cells. Furthermore, a new statistical methodology quantifies the progression of normal cells into abnormal cells.

\vspace{-3mm}
\subsection{Key Contributions}

The major contributions of this article are as follows:

\begin{enumerate}

\item A comprehensive framework encompassing three distinct stages: the segmentation of cellular components, the classification of the cell, and the evaluation of RA.

\item A DL architecture, Multi-Resolution Fusion in Deep Convolutional Network (MRF-DCN) for cervical cell classification consists of a significantly lower number of learnable parameters than the state-of-the-art models. This architecture processes images at varying scales and aspect ratios.

\item A novel DL model that combines Multi-Task Learning (MTL) to complete segmentation and classification tasks at the same time.

\item A statistical method for RA by leveraging the features extracted from images using the trained model, enabling a probabilistic approach to predict the likelihood of progression from normal cells to abnormal cells. 

\end{enumerate} 

\vspace{-4mm}
\subsection{End-to-End Framework}

\begin{figure*}[!htp]
\centering
\includegraphics[scale=0.4]{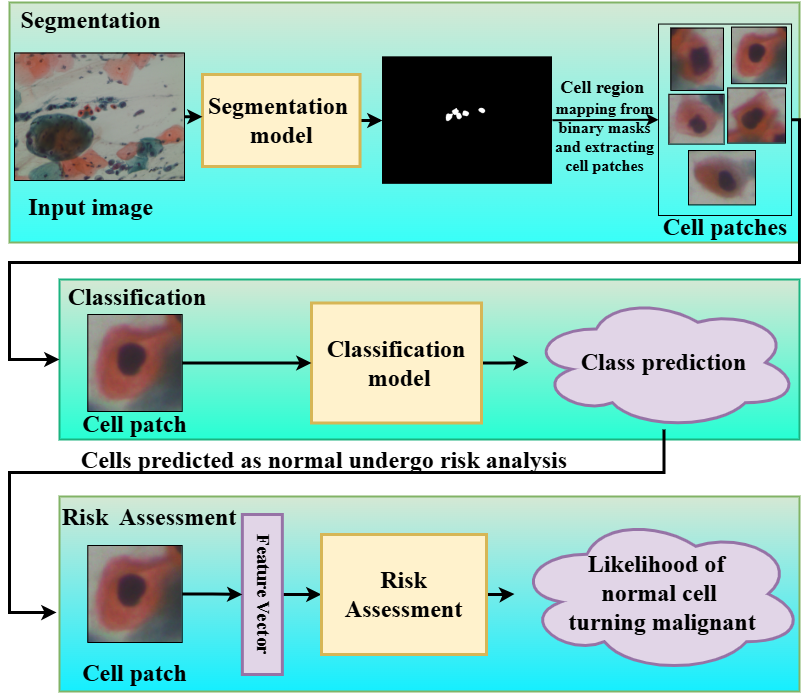}
\caption{ Proposed framework for cervical cancer diagnosis}
\label{fig.pipeline}
\end{figure*}

The proposed framework architecture shown in Fig. \ref{fig.pipeline} comprises three main stages: segmentation, classification, and RA. Initially, cells are segmented from whole slide images using the SIPaKMeD dataset. For this, bounding boxes are drawn around the segmented regions to extract the Regions of Interest (ROIs). These extracted regions are passed to a classification model to determine the type of the cells. Furthermore, an MTL technique performs segmentation and classification simultaneously, streamlining the process. In the final stage, after classifying the cell, a 64-dimensional feature vector is extracted to assess the likelihood of normal cells progressing to abnormal states. This comprehensive framework ensures precise segmentation, accurate classification, and effective RA. The remaining article is organized as follows. Section \ref{Datasets-section} describes the datasets employed in the study. Section \ref{methodology} presents the methodology of the proposed models. Section \ref{results} explains the results and discussions. Section \ref{conclusion} concludes the study.

\vspace{-3mm}
\section{Datasets and Pre-Processing} \label{Datasets-section}

\begin{figure*}[htbp]
    \centering
    \subfloat[Dyskaryotic]{
        \includegraphics[scale=0.060]{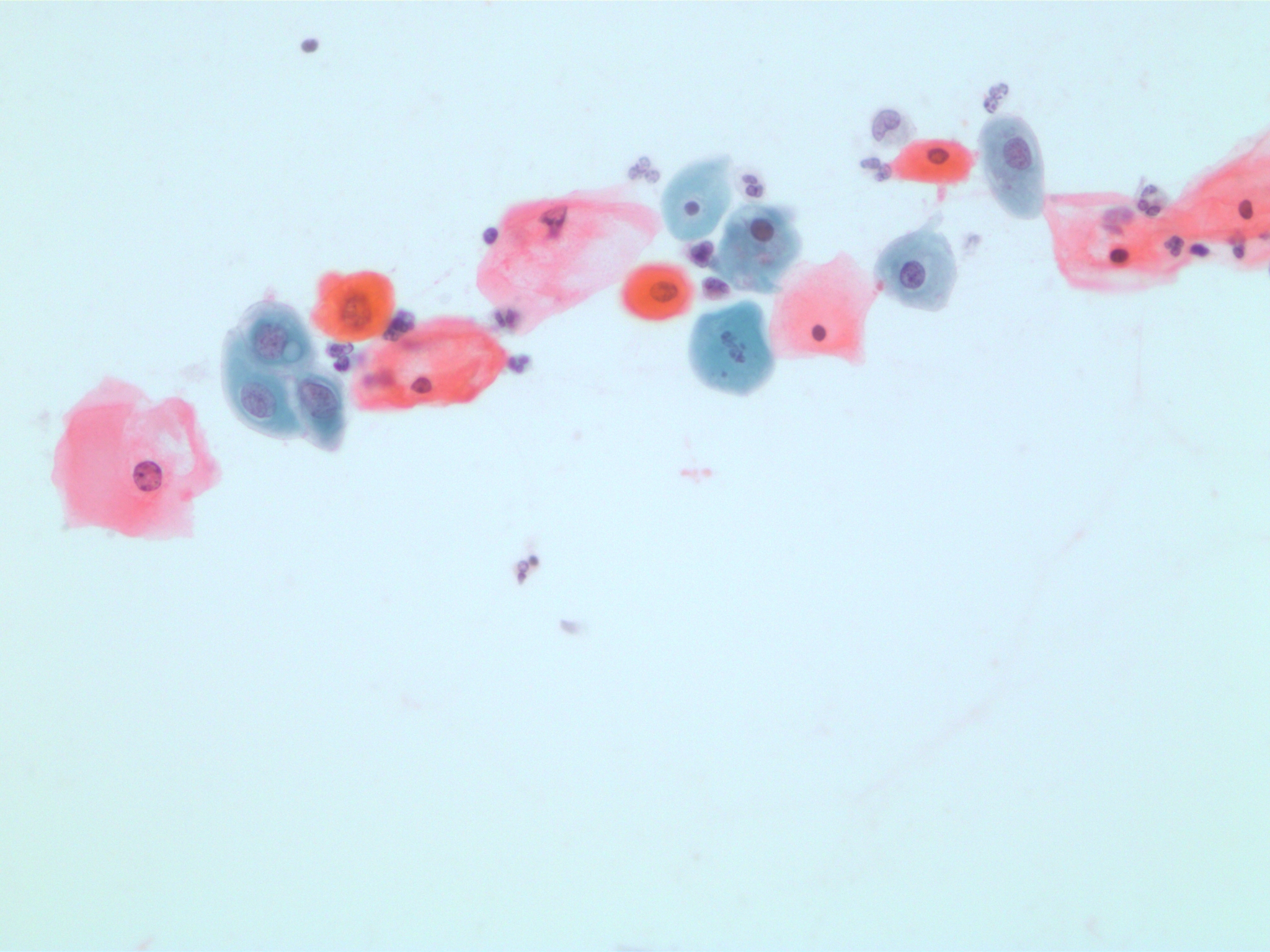}
        \label{fig:dyk_sample}
    }
    \subfloat[Koilocytotic]{
        \includegraphics[scale=0.045]{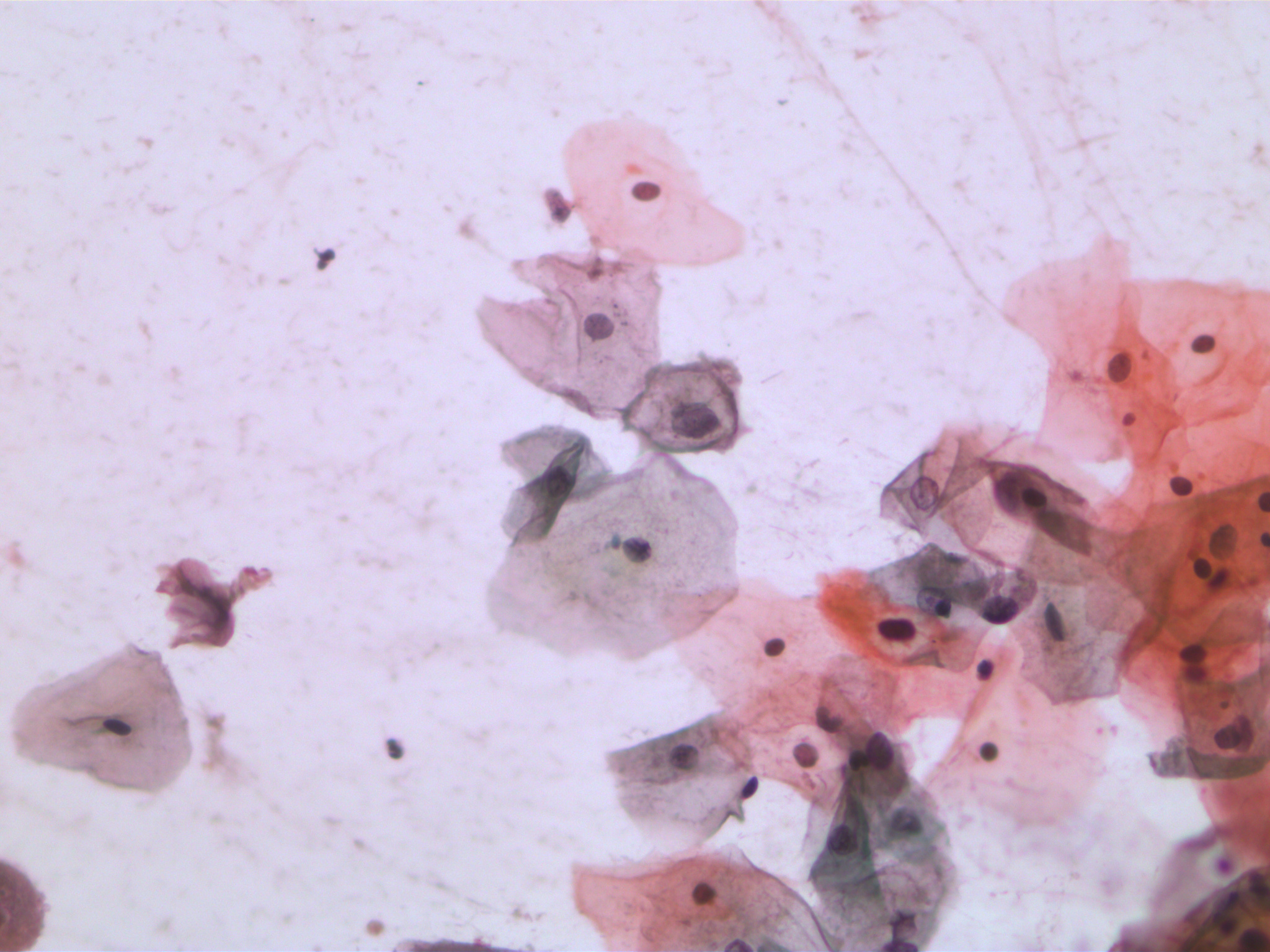}
        \label{fig:koil_sample}
    }
    \subfloat[Metaplastic]{
        \includegraphics[scale=0.045]{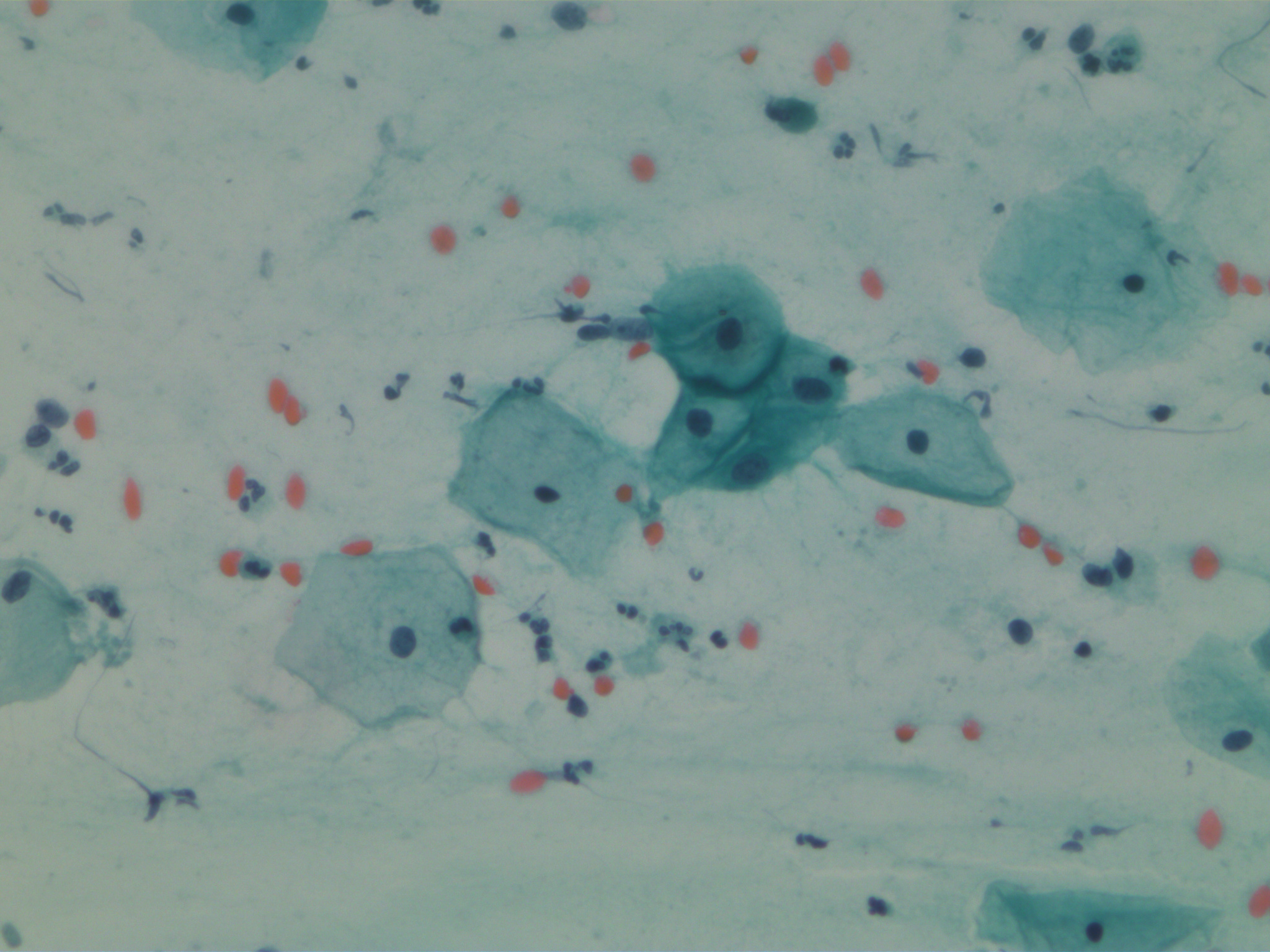}
        \label{fig:meta_sample}
    }
    \subfloat[Parabasal]{
        \includegraphics[scale=0.045]{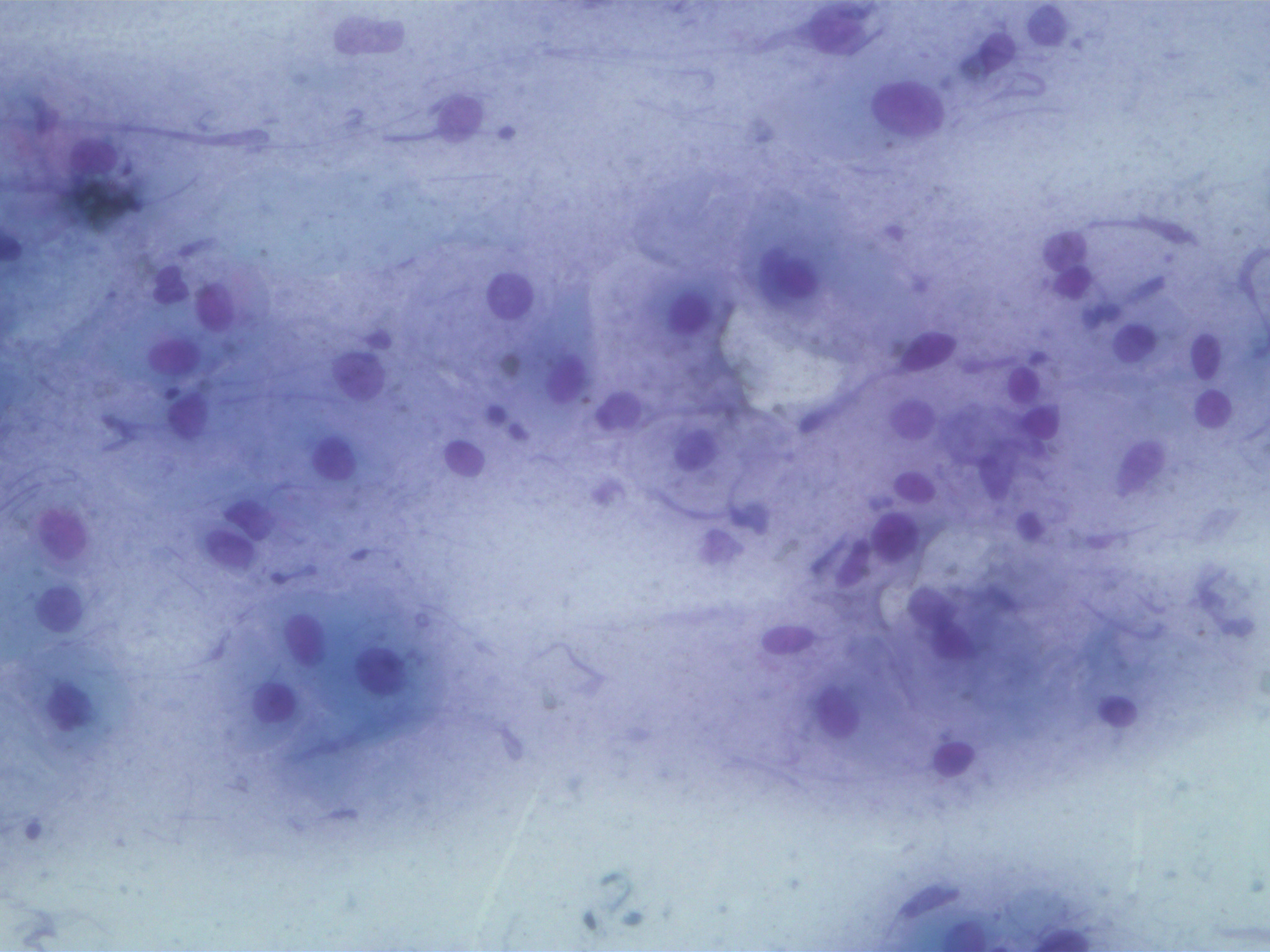}
        \label{fig:para_sample}
    }
    \subfloat[Superficial-Intermediate]{
        \includegraphics[scale=0.045]{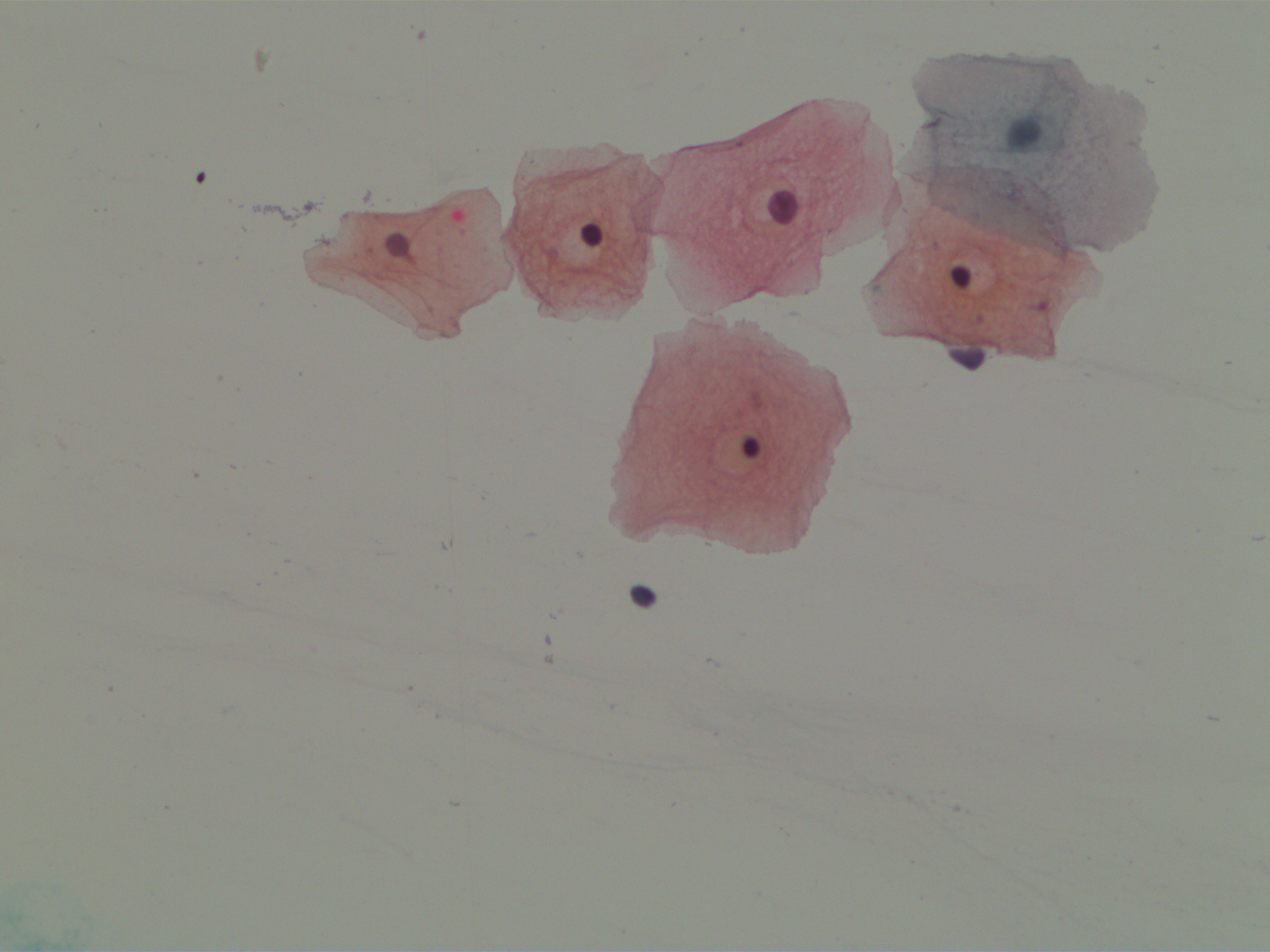}
        \label{fig:super_sample}
    }
    \caption{Examples of WSI}
    \label{fig:sipakmed_wsi}
\end{figure*}

\begin{figure}[]
    \centering
    \subfloat[]{
        \includegraphics[width=2cm,height=2cm]{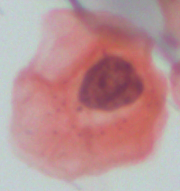}
        \label{fig:dyk_cell}
    }
    \subfloat[]{
        \includegraphics[width=2cm,height=2cm]{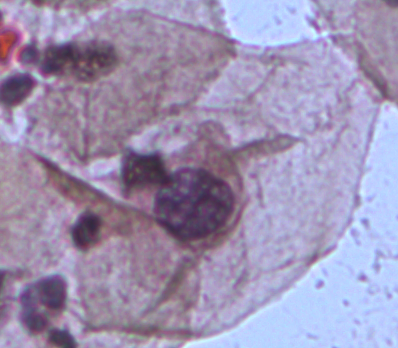}
        \label{fig:koil_cell}
    }
    \subfloat[]{
        \includegraphics[width=2cm,height=2cm]{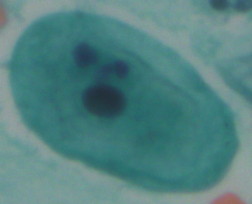}
        \label{fig:meta_cell}
    }
    \hfill
    \subfloat[]{
        \includegraphics[width=2cm,height=2cm]{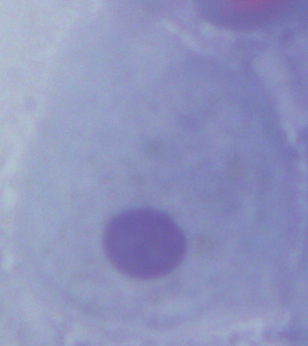}
        \label{fig:para_cell}
    }
    \subfloat[]{
        \includegraphics[width=2.5cm,height=2cm]{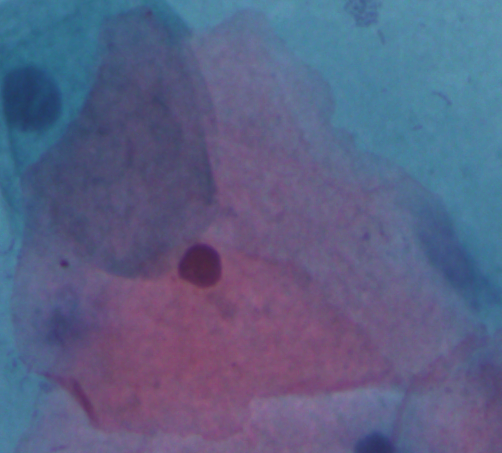}
        \label{fig:super_cell}
    }
    \caption{Examples of cell patches: (a) Dyskaryotic (b) Koilocytotic (c) Metaplastic (d) Parabasal (e) Superficial-Intermediate }
    \label{fig:sipakmed_cells}
\end{figure}

In this study, we used the SIPaKMeD dataset \cite{sipakmeddataset}. The SIPaKMeD database comprises a total of 4049 images of individual cells that were manually isolated from 966 high-resolution cluster cell images with a resolution of 2048x1536, obtained from Pap smear slides, which are glass slides holding samples of cervical cells collected during a Pap smear test. A charge-coupled device camera mounted on an optical microscope was used to take these images. The cell images are categorized into five groups. Here are some Whole Slide Images (WSI) in Fig. \ref{fig:sipakmed_wsi} and cell images in  Fig. \ref{fig:sipakmed_cells} are provided as examples. The Table \ref{table:2} displays the number of images in each class.

\begin{table}[htp]
    \caption{SIPaKMeD Dataset Distribution}
    \label{table:2}
    \setlength{\tabcolsep}{3pt}
    \centering
    \resizebox{\columnwidth}{!}{%
    \begin{tabular}{|c|c|c|c|c|}
        \hline
        \textbf{Class} & \textbf{Type of Cancer Cell} & \textbf{Category} & \textbf{WSI} & \textbf{Cropped Cells} \\
        \hline
        1 & Superficial-Intermediate & Normal & 126 & 831 \\
        2 & Parabasal & Normal & 108 & 787 \\
        3 & Koilocytotic & Abnormal & 238 & 825 \\
        4 & Dyskaryotic & Abnormal & 271 & 793 \\
        5 & Metaplastic & Benign & 223 & 813 \\
        \hline
        \multicolumn{3}{|c|}{\textbf{Total}} & \textbf{966} & \textbf{4049} \\
        \hline
    \end{tabular}%
    }
\end{table}

\vspace{-2mm}

\vspace{-6mm}

\section{Proposed Framework} \label{methodology}

DL-based cervical cancer diagnosis is a contemporary active field with significant potential to surpass the statistical models. Similarly, in the current study, two robust DL models and one statistical method for RA are introduced that remain uninvestigated for cervical cancer diagnosis. The first model is designed for classification, while the second DL model leverages MTL to simultaneously perform segmentation and classification, offering a more comprehensive approach to cervical cancer diagnosis. 

\vspace{-4mm}

\subsection{Segmentation}

We utilized the WSI images from the SIPaKMeD dataset for the segmentation task, where each high-resolution image of size $2048\times1536$ was divided into 20 overlapping patches of size $512\times512$. Binary masks were generated for each patch, where white pixels represent the cell regions, and black pixels denote the background. To enhance the dataset's relevance and eliminate redundant data, we filtered out patches containing no cell regions identified as black images. The segmentation stage focuses on delineating the cell boundaries in the images from the SIPaKMeD dataset. This is approached as a binary segmentation task, which aims to differentiate the cell regions from the background. To achieve this, four DL models named UNet \cite{unet-seg}, LinkNet \cite{linknet-seg}, FPN \cite{fpn-seg}, and PSPNet \cite{pspnet-seg} were trained to generate precise segmentation masks. These models were implemented using a publicly available library of segmentation models \cite{githubsm}, enabling the application of state-of-the-art architectures.

Further, we employed a method to draw bounding boxes around segmented cells identified through binary segmentation. The binary segmentation output consisted of a mask where white pixels represented the segmented cell regions, and black pixels denoted the background. To localize these regions, we extracted the coordinates \((x, y)\) of all white pixels. The bounding box was defined by calculating the minimum and maximum pixel values along both axes: $x_{\text{min}} = \min(x)$, $x_{\text{max}} = \max(x)$, $y_{\text{min}} = \min(y)$, $y_{\text{max}} = \max(y)$. Since the segmentation mask resolution was identical to the original image resolution, no scaling of these coordinates was required. To ensure the rectangle encompassed the entire cell region without truncation, a padding offset \((O)\) was added, resulting in the final coordinates: $[x_{\text{min}} - O, \, y_{\text{min}} - O, \, x_{\text{max}} + O, \, y_{\text{max}} + O]$. These coordinates define the top-left \((x_{\text{min}} - O, y_{\text{min}} - O)\) and bottom-right \((x_{\text{max}} + O, y_{\text{max}} + O)\) corners of the rectangle, which were then used to draw a bounding box directly on the original image, providing a precise visual representation of the segmented cells and aiding in further analysis. This stage forms the foundation for subsequent steps in the pipeline by accurately isolating the regions of interest for further analysis. 

\vspace{-3mm}
\subsection{MRF-DCN}
\begin{figure*}[hbt!]
\centering
\includegraphics[scale=0.19]{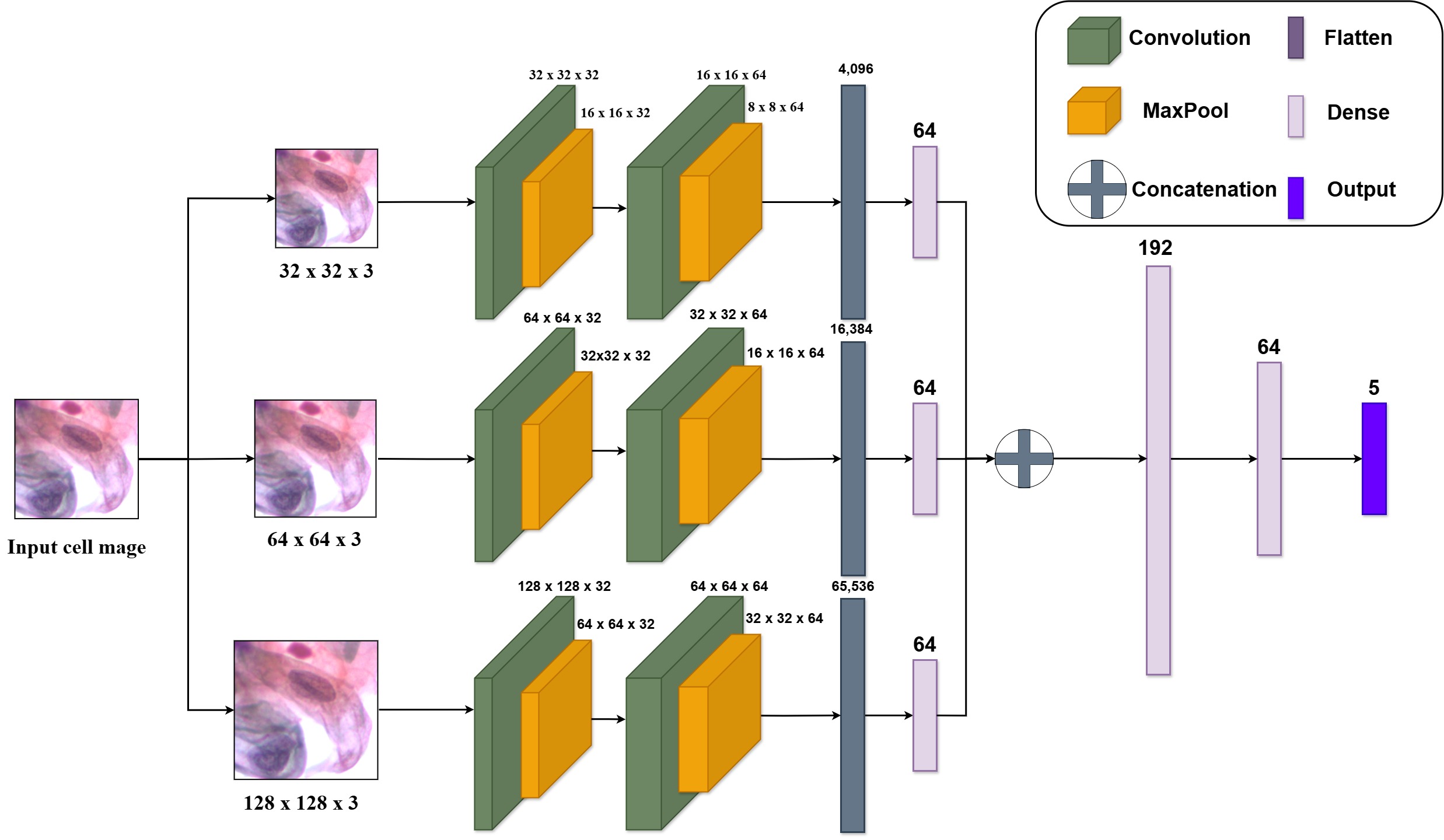} 
\caption{ Multi-Resolution Fusion in Deep Convolutional Network (MRF-DCN)}
\label{fig.mrcnn_bd}
\end{figure*}

After completing the segmentation task, which involved extracting the location of cancer cells in WSI, we proceeded to the classification stage in our proposed framework. In this stage, we developed a lightweight DL model named MRF-DCN, shown in Fig. \ref{fig.mrcnn_bd}. Unlike traditional deep architectures, MRF-DCN employs only a single multi-resolution feature extraction module. While this design deviates from the conventional "deeper is better" philosophy, it was intentionally crafted to balance computational efficiency and robust feature representation. Despite its shallow structure, MRF-DCN performs comparably to state-of-the-art models due to its ability to process multiple spatial resolutions in parallel, effectively capturing coarse and fine-grained features. The multi-resolution input strategy contributes to better generalization and acts as an implicit regularizer, particularly beneficial in relatively limited and imbalanced datasets. Furthermore, using depthwise separable convolutions reduces the number of learnable parameters while preserving spatial and contextual information across scales.

For a fair comparison, we trained several standard DL models from scratch, including  Alexnet \cite{alexnet}, DenseNet121 \cite{densenet121}, ResNet18 and ResNet101~\cite{resnet}, MobileNetV3-Large \cite{mobilenetv3-large-1}, and VGG19 \cite{vgg19}, using the same training data and settings. However, these models are architecturally complex.

Overall, MRF-DCN achieves a substantial trade-off between performance and architectural simplicity. It is suitable for deployment in computationally constrained clinical environments while delivering classification accuracy on par with more profound, complex networks.

Traditional methods often resize all images to a fixed size, which can lead to interpolation or compression, resulting in significant information loss. To mitigate this issue, we designed a model accommodating input images of varying sizes. By allowing the model to analyze multiple copies of the same image in different dimensions, we preserve the intrinsic details and unique features of each image. This architecture features three branches, each taking images as input in various aspect ratios: 32x32, 64x64, and 128x128. Our architecture delivered comparable performance with significantly fewer learnable parameters than existing models. This efficiency underscores the effectiveness of our design in balancing complexity and accuracy. Each branch of the MRF-DCN processes images at different resolutions and follows a consistent structure. It starts with two convolutional layers featuring 32 and 64 filters, respectively, which use 3x3 kernels, a stride of 1, and padding of 1 to maintain the spatial dimensions. A max-pooling layer with a 2x2 kernel and a stride of 2 follows, reducing the spatial dimensions by half. Finally, the feature maps are flattened, and a fully connected layer reduces the dimensionality to 64 features. Each branch produces a 64-dimensional output, and the outputs from the three branches are concatenated to form a combined feature vector of 192 dimensions. This combined feature vector is then passed through another fully connected layer to reduce the dimensionality to 64 features. Finally, the output layer computes the class probabilities using the softmax activation function.

\vspace{-3mm}
\subsection{Multi-Task UNet with a Squeezed Bottomneck Layer}
\vspace{-1mm}

\begin{figure*}[hbt!]
\centering
\includegraphics[scale=0.35]{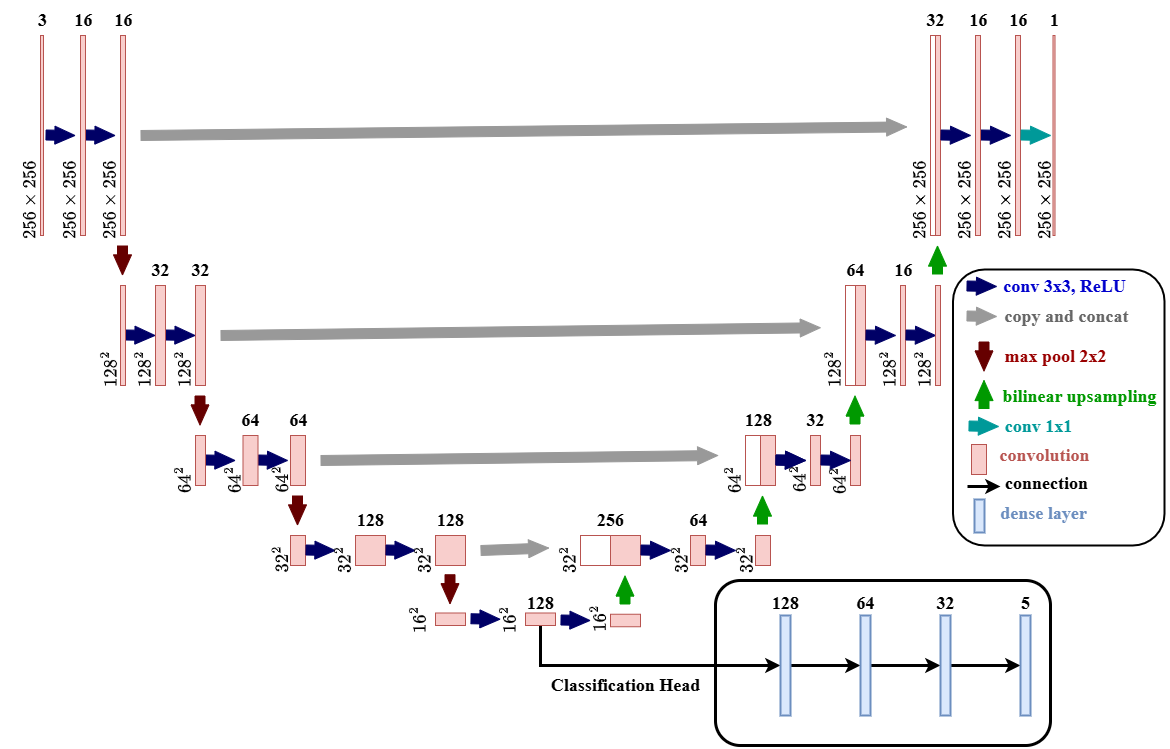} 
\caption{Multi-task UNet with a squeezed bottomnecklayer}
\label{fig:mtl_unet}
\end{figure*}

To address the dual objectives of segmentation and classification, we implemented a Multi-Task Learning (MTL) framework by modifying the standard UNet architecture, resulting in a model we refer to as the “Multi-task UNet with a squeezed bottleneck layer,” as illustrated in Fig. \ref{fig:mtl_unet}. In this design, the bottleneck layer of UNet was compressed and extended with a sequential layer dedicated to the classification task. During training, two separate loss functions—segmentation loss and classification loss—were computed and combined to update the network weights through backpropagation jointly. This MTL approach offers several advantages: it enables the model to leverage shared representations, thereby improving generalization across both tasks; it reduces computational overhead by eliminating the need for separate models; and it facilitates consistency between the segmentation and classification outputs, which is particularly beneficial in medical image analysis where spatial and categorical information are often correlated. By jointly optimizing both tasks, the model learns more robust and task-aware features, ultimately enhancing overall performance and efficiency.

\vspace{-3mm}
\subsection{Risk Assessment}
\begin{figure}[hbt!]
\centering
\includegraphics[scale=0.18]{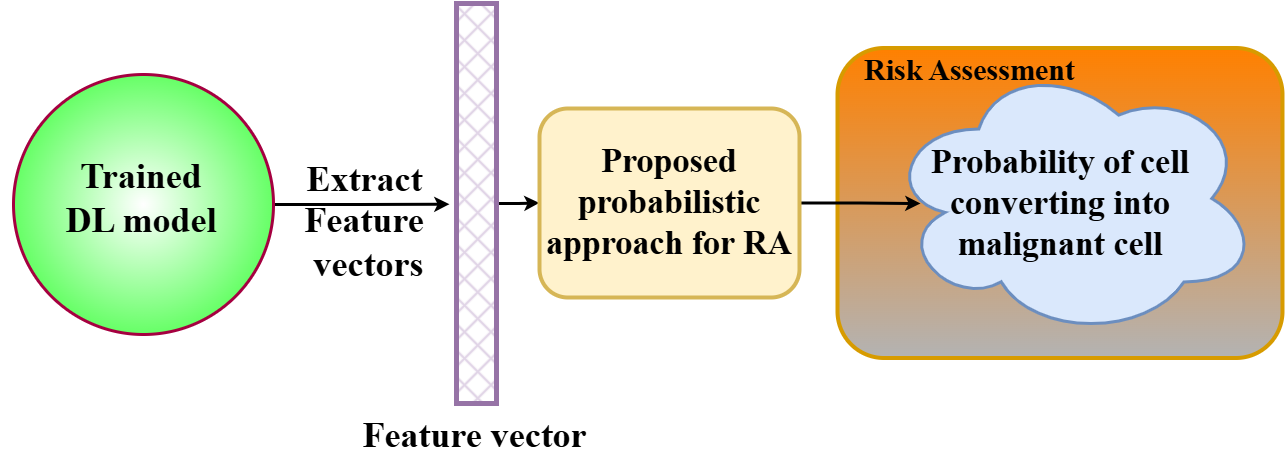} 
\caption{ Proposed Approach for Risk Assessment}
\label{fig.ra_app}
\end{figure}

The central aim of this approach lies in estimating the likelihood of normal cells transforming abnormal cells. This is facilitated by extracting a 64-dimensional feature vector from each cell image, representing the cell's characteristics. Leveraging this feature vector, a probabilistic score is computed to quantify the conversion potential, as shown in Fig \ref{fig.ra_app}. A comprehensive elucidation of the methodology is presented in the subsequent subsections.

\subsubsection{Feature Vector Extraction}

Our proposed method aims to predict the probability of future superficial-intermediate, parabasal, and metaplastic converting to koilocytotic and dyskeratotic cells. We achieved this by training the entire SIPaKMeD dataset using our proposed model, MRF-DCN. After training the model, we extracted a feature vector of length 64 for each image from the last layer before the final output layer. This transformation allowed us to represent cell image information in a compact vector format. Subsequently, we used these feature vectors to train several standard ML models, including Logistic Regression (LR), KNN, Naive Bayes, SVM, SVM with Radial Basis Function (RBF) kernel, RF, and Decision Tree (DT). These models were employed to determine whether the extracted features could effectively distinguish between the four classes. To assess the risk of converting normal to abnormal cells, we computed the mean, variance and covariance of the feature vectors for each class. For any test image, we extracted its feature vector and measured the distance between this vector and the means of the classes. Using this distance, we calculated the probability of the test image belonging to each class. The detailed procedure and mathematical model are provided in the subsequent section.

\subsubsection{Probabilistic Approach for calculating RA}
In this section, we describe a probabilistic method for classifying an unseen sample into one of multiple classes based on pre-calculated mean and covariance matrices. Additionally, we provided a measure of the sample's similarity to the remaining classes. For each class $i$ in the dataset, we first calculated the mean vector $\mu_i$ and the covariance matrix $\Sigma_i$ for each class $i$. The unseen sample $x$ is then classified using the multivariate Gaussian distribution. The probability density function of the multivariate Gaussian distribution for class $i$ is defined in Eq. \ref{gaussian_equation}.

\begin{equation} \label{gaussian_equation}
P(x|\mu_i, \Sigma_i) = \frac{1}{\sqrt{(2\pi)^d |\Sigma_i|}} \exp\left(-\frac{1}{2}(x - \mu_i)^T \Sigma_i^{-1} (x - \mu_i)\right)
\end{equation}

Where:
\begin{itemize}
    \item $d$: is the dimensionality of the feature vector
    \item $|\Sigma_i|$ : is the determinant of the covariance matrix
    \item $\Sigma_i^{-1}$ : is the inverse of the covariance matrix
\end{itemize}

To assign the sample $x$ to a class, we apply Bayes' theorem to compute the posterior probability $P(C_i|x)$ for each class $C_i$. Bayes' theorem is given in Eq. \ref{bayes_equation}, where $P(C_i|x)$ is the posterior probability of class $C_i$ given the sample $x$, $P(x|C_i)$ is the likelihood of the sample $x$ under the class $C_i$, $P(C_i)$ is the prior probability of class $C_i$, and $P(x)$ is the evidence or the total probability of the sample $x$ under all classes. Assuming equal prior probabilities $P(C_i)$ for all classes, the priors cancel out, and we have the following Eq. \ref{prop_equat}.

\begin{equation} \label{bayes_equation}
P(C_i|x) = \frac{P(x|C_i) P(C_i)}{P(x)}
\end{equation}

\begin{equation} \label{prop_equat}
P(C_i|x) \propto P(x|C_i)
\end{equation}

The class into which the sample $x$ is assigned has the highest posterior probability $P(C_i|x)$ as shown in Equation \ref{argmax}, where  $\arg\max$ function identifies the index $i$ of the class $C_i$ that maximizes the posterior probability $P(C_i|x)$. It determines the most probable class for the sample $x$ by selecting the class with the highest probability among all possible classes. To measure the similarity of the sample $x$ to the remaining classes, we normalize the likelihoods to obtain posterior probabilities that sum to one. The normalization is done by dividing each likelihood $P(x|C_i)$ by the total probability $P(x)$ as described in Eq. \ref{normalization_eq}.

\begin{equation} \label{argmax}
\text{Class} = \arg\max_{i} P(C_i|x)
\end{equation}

\begin{equation} \label{normalization_eq}
P_n(C_i|x) = \frac{P(x|C_i)}{\sum_{j} P(x|C_j)}
\end{equation}

\begin{algorithm}
\caption{Risk Assessment}
\label{alg:RA}
\SetAlgoLined
\DontPrintSemicolon
\KwIn{Classes $\{C_1, C_2, \dots, C_n\}$ with feature vectors, test sample $x$}
\KwOut{Predicted class $\hat{x}$, $P_n(C_i|x)$, and cosine similarity}
\For{$C_i$ in classes $\{C_1, C_2, \dots, C_n\}$}{
     Compute $\mu_i$, $\Sigma_i$ for class $C_i$\;
    Compute $P(x|\mu_i, \Sigma_i)$ using Eq.~\ref{gaussian_equation}\;
    Compute $P(C_i|x)$ using Eq.~\ref{bayes_equation}\;
}
Normalize $P(C_i|x)$ using Eq.~\ref{normalization_eq}\;
Predict $\hat{x}$ using the $\arg\max$ function using Eq.~\ref{argmax}\;
\For{$C_i$ in classes $\{C_1, C_2, \dots, C_n\}$}{
    Compute cosine similarity between $x$ and $\mu_i$ using Eq.~\ref{cosine_formula}\;
}
\Return $\hat{x}$, $P_n(C_i|x)$, and cosine similarity
\end{algorithm}

These posterior probabilities quantitatively measure how similar the sample is to each class. The predicted class has the highest posterior probability, whilst their posterior probabilities indicate the likelihood or resemblance of the other classes. Additionally, we measured cosine similarity to check the similarity of the sample to the classes. The cosine distance between two vectors $a$ and $b$ is given by Equation \ref{cosine_formula}.

\begin{equation} \label{cosine_formula}
\text{cosine similarity}(a, b) =  \frac{a \cdot b}{|a| |b|}
\end{equation}

where $a \cdot b$ is the dot product of vectors $a$ and $b$, and $|a|$ and $|b|$ are the magnitudes of vectors $a$ and $b$, respectively. Cosine similarity was an effective measure of similarity and worked well in RA, providing an additional perspective on the closeness of the sample to each class. The algorithm \ref{alg:RA} outlines the entire RA method. By employing this probabilistic approach and incorporating cosine similarity, we can effectively classify the unseen sample and quantify its similarity to all classes, providing a comprehensive classification and similarity assessment method.

In summary, our developed method offers a predictive approach to assess the probability of normal cells progressing to abnormal cells. Furthermore, this approach is effective on the Mendeley LBC Dataset \cite{mendleylbc} as well, enabling the assessment of the probability of low-grade squamous intraepithelial lesion and high-grade squamous intraepithelial lesion progressing to squamous cell carcinoma. 

\vspace{-4mm}
\section{Results and Discussions} \label{results}

This section presents the detailed simulation results of the proposed study. All computations were implemented in Python, and all experiments used a Nvidia A100 80 Gb GPU graphics card and Linux operating system.
\vspace{-2mm}
\subsection{Performance Measures}
This study uses two kinds of performance measurements: segmentation and classification. The segmentation performance is summarized using the Intersection of Union (IoU), while the classification performance is evaluated based on F1 score, accuracy, recall, and precision. Eq. \ref{form_iou}-\ref{form_f1} describe the formulae for performance metrics.

\begin{equation} \label{form_iou}
    \text{IoU} = \frac{\text{Area of Overlap}}{\text{Area of Union}} = \frac{TP}{TP + FP + FN}
\end{equation}

\begin{equation} \label{form_acc}
\text{Accuracy} = \frac{TP + TN}{TP + TN + FP + FN}
\end{equation}

\begin{equation} \label{form_prec}
\text{Precision} = \frac{TP}{TP + FP}
\end{equation}

\begin{equation} \label{form_rec}
\text{Recall} = \frac{TP}{TP + FN}
\end{equation}

\begin{equation} \label{form_f1}
\text{F1-score} = 2 \cdot \frac{\text{Precision} \cdot \text{Recall}}{\text{Precision} + \text{Recall}}
\end{equation}

Here, $TP$ represents True Positives, $TN$ represents True Negatives, $FP$ represents False Positives, and $FN$ represents False Negatives.

\vspace{-2mm}
\subsection{Segmentation}
The current subsection presents the binary segmentation of the cell, where the primary objective is to distinguish the cell regions from the background. As previously discussed, four DL models—UNet, LinkNet, FPN, and PSPNet—were trained to generate segmentation masks. Contrary to earlier expectations, UNet demonstrated the highest performance across most evaluation metrics, achieving the highest test IoU score of 0.81, a DC of 0.68, and an overall accuracy of 0.97. PSPNet also showed competitive performance with a test IoU of 0.77 and matched UNet in DC which is 0.68 while maintaining a high accuracy of 0.96. FPN achieved slightly lower performance with an IoU of 0.75 and DC of 0.63. LinkNet, despite attaining a high training IoU of 0.97, underperformed during testing, recording a significantly lower test IoU of 0.55 and DC of 0.35, suggesting overfitting. These comparative results, including class-wise precision, recall, and F1-scores, are systematically summarized in Table \ref{table:seg_results}, highlighting UNet and PSPNet as the most robust architectures for binary segmentation in this context.

\begin{table*}[hbt!]

\caption{Comparison of DL models for binary cell segmentation using IoU, DC, Accuracy, and class-wise metrics}
\centering

\begin{tabular}{|l|c|c|c|c|c|c|c|c|c|c|}
\hline
\textbf{Models} & \textbf{IoU} & \textbf{DC} & \textbf{Accuracy} & \multicolumn{2}{c|}{\textbf{Precision}} & \multicolumn{2}{c|}{\textbf{Recall}} & \multicolumn{2}{c|}{\textbf{F1-Score}} & \textbf{Train IoU} \\ \cline{5-10}
 & & & & \textbf{0} & \textbf{1} & \textbf{0} & \textbf{1} & \textbf{0} & \textbf{1} & \\ \hline
UNET \cite{unet-seg}       & 0.81 & 0.68 & 0.97 & 0.98 & 0.87 & 0.98 & 0.85 & 0.98 & 0.86 & 0.96 \\ \hline
LinkNET \cite{linknet-seg} & 0.55 & 0.35 & 0.88 & 0.94 & 0.45 & 0.92 & 0.50 & 0.93 & 0.47 & 0.62 \\ \hline
FPN \cite{fpn-seg}         & 0.75 & 0.63 & 0.96 & 0.97 & 0.86 & 0.98 & 0.77 & 0.98 & 0.81 & 0.76 \\ \hline
PSPNet \cite{pspnet-seg}   & 0.77 & 0.68 & 0.96 & 0.98 & 0.85 & 0.98 & 0.80 & 0.98 & 0.82 & 0.92 \\ \hline
\end{tabular}
\label{table:seg_results}
\end{table*}

\subsection{Classification}
\begin{figure}[hbt!]
\centering
\includegraphics[scale=0.45]{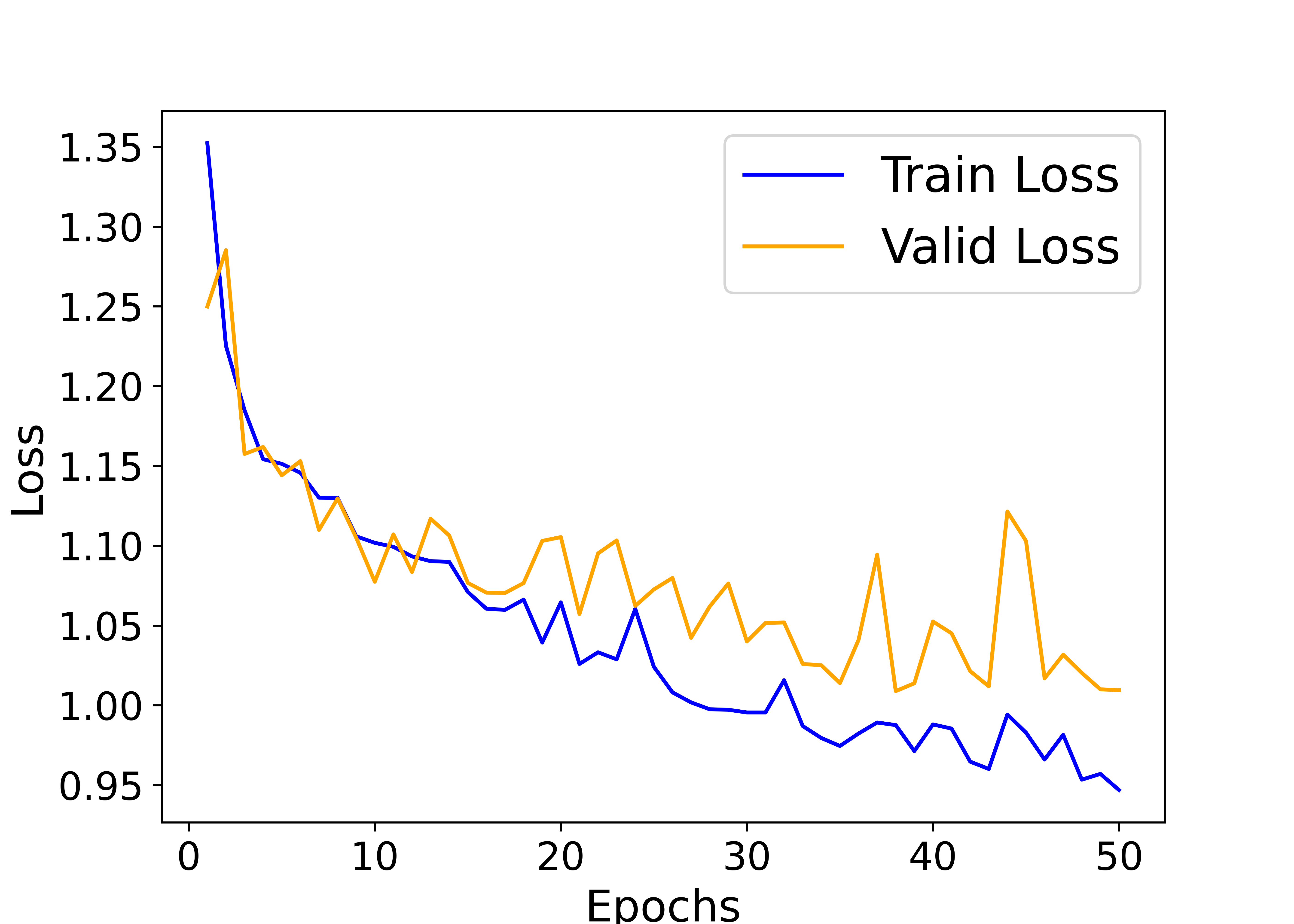} 
\caption{MRF-DCN: CE Loss}
\label{fig:mrfdcn_loss}
\end{figure}

\begin{figure}[hbt!]
\centering
\includegraphics[scale=0.45]{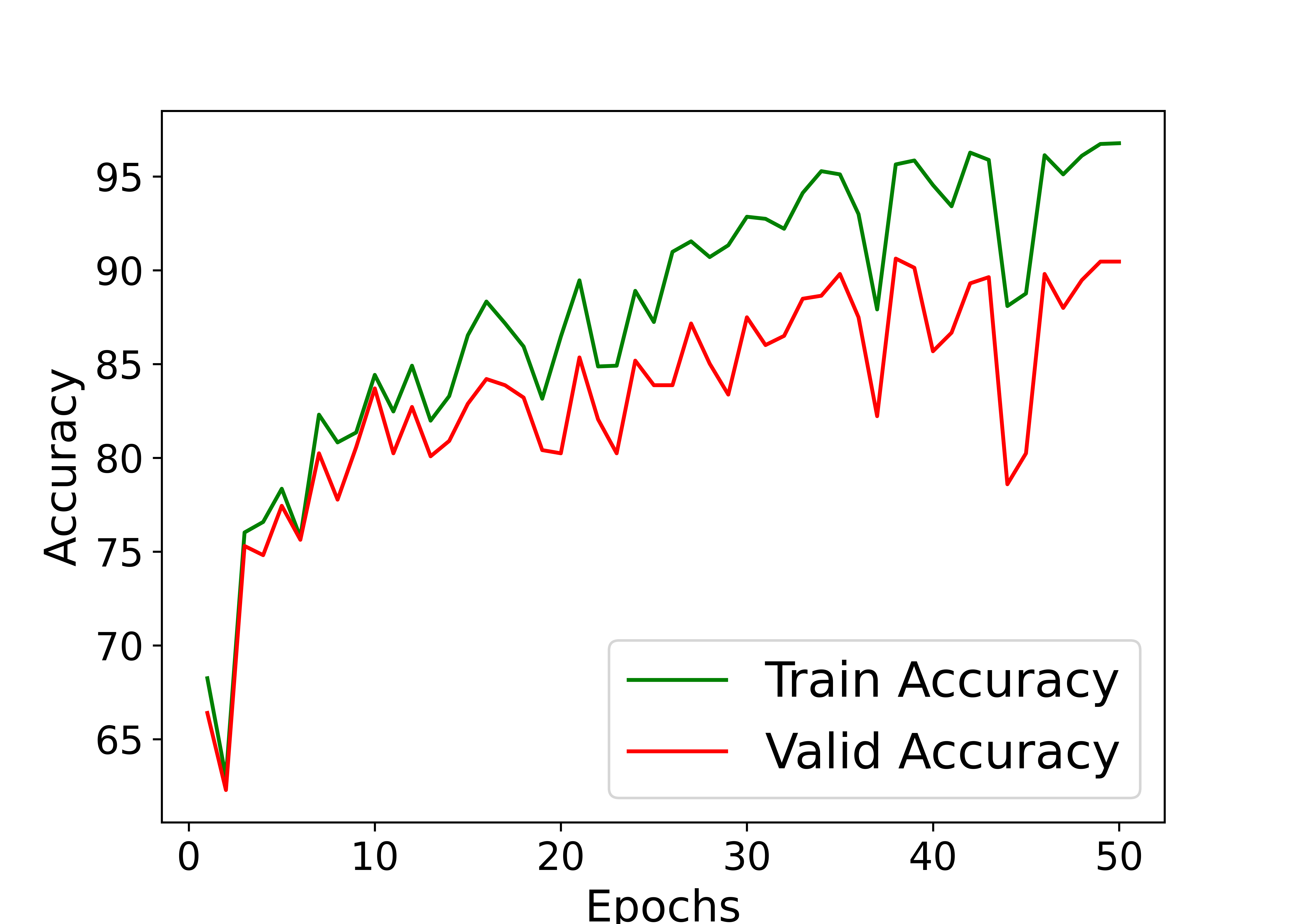} 
\caption{MRF-DCN Accuracy}
\label{fig:mrfdcn_acc}
\end{figure}

The current subsection stages the study implementing the MRF-DCN model. Fig. \ref{fig:mrfdcn_loss} and \ref{fig:mrfdcn_acc} illustrate the training process of our MRF-DCN model. The model's hyperparameters for training are detailed in the Table \ref{tab:mrf-dcn-config}. The proposed model's performance was meticulously evaluated and demonstrated in Table \ref{tab:mrf-dcn-result}. 

\begin{table}[htp]
    \caption{MRF-DCN Hyperparameter setting values}
    \label{tab:mrf-dcn-config}
    \centering
    \resizebox{\columnwidth}{!}{%
    \begin{tabular}{|c|c|}
        \hline
        \textbf{Hyperparameter} & \textbf{Value} \\
        \hline
        Train-test-valid split ratio & 70\% - 15\% - 15\% \\ 
        Optimizer & Adam \cite{adam_optimizer} \\ 
        Loss & Cross Entropy (CE) \\ 
        Learning rate & 0.001 \\ 
        Batch size & 32 \\ 
        Epochs & 50 \\ 
        Learnable parameters & 1.7 Million \\ 
        \hline
    \end{tabular}%
    }
\end{table}

\begin{table}[htp]
    \caption{Performance of MRF-DCN}
    \label{tab:mrf-dcn-result}
    \centering
    \resizebox{\columnwidth}{!}{%
    \begin{tabular}{|c|c|c|c|}
        \hline
        \textbf{Class} & \textbf{Precision} & \textbf{Recall} & \textbf{F1-Score} \\ 
        \hline
        Metaplastic & 0.88 & 0.88 & 0.88 \\ 
        Superficial-Intermediate & 0.95 & 0.91 & 0.93 \\ 
        Parabasal & 0.98 & 0.93 & 0.96 \\ 
        Koilocytotic & 0.80 & 0.84 & 0.82 \\ 
        Dyskeratotic & 0.91 & 0.96 & 0.93 \\ \hline
        \textbf{Average} & 0.90 & 0.90 & 0.90 \\ 
        \hline
    \end{tabular}%
    }
\end{table}

\begin{figure}
\centering
\includegraphics[scale=0.56]{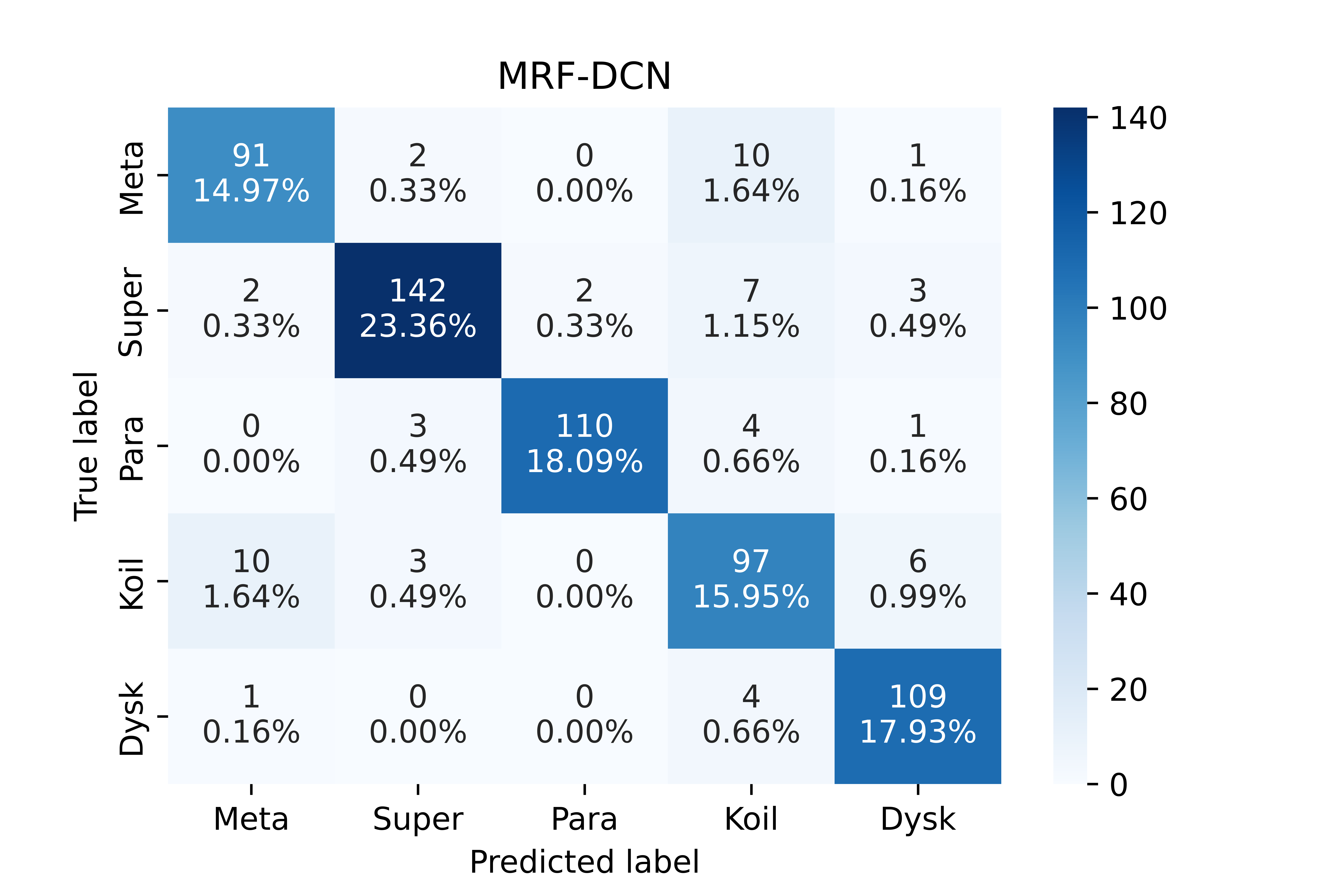}
\caption{ Confusion matrix illustrating the performance of MRF-DCN }
\label{fig.cf_mrfdcn}
\end{figure}

Fig. \ref{fig.cf_mrfdcn} details the confusion matrix of the testing dataset. The model demonstrates strong classification performance for the superficial-intermediate class, correctly identifying 142 instances, although occasional confusion exists with the koilocytotic and dyskaryotic classes. The metaplastic and koilocytotic classes exhibit higher misclassification rates, with koilocytotic often confused with metaplastic, indicating a potential overlap in features between these classes. The dyskaryotic class is accurately classified in 109 cases but is sometimes misidentified as koilocytotic or metaplastic. On the other hand, the parabasal class shows relatively lower misclassification rates than metaplastic and koilocytotic, suggesting that the model is better at distinguishing parabasal from the different classes. These observations highlight the model's effectiveness in certain areas while indicating specific class pairs, such as metaplastic and koilocytotic, where improvements could be made to reduce misclassification.

\begin{table}[htp]
    \caption{Performance comparison of the MRF-DCN model with existing models}
    \label{tab:comp_mrcnn}
    \centering
    \resizebox{\columnwidth}{!}{%
    \begin{tabular}{|c|c|c|c|c|c|}
        \hline
        \textbf{DL models} & \textbf{Parameters} & \textbf{Accuracy} & \textbf{Precision} & \textbf{Recall} & \textbf{F1-Score} \\ 
        \hline
        \textbf{MRF-DCN}    & \textbf{1.7 M}  & \textbf{91.28}  & \textbf{0.91} & \textbf{0.91} & \textbf{0.91} \\ \hline
        Mobilenet-v3        & 5.47 M          & 89.11           & 0.90          & 0.88          & 0.88          \\ \hline
        Densenet-121        & 8 M             & 89.38           & 0.90          & 0.89          & 0.89          \\ \hline
        ResNet-18           & 11.69 M         & 85.68           & 0.86          & 0.86          & 0.85          \\ \hline
        ResNet-50           & 25.6 M          & 87.30           & 0.88          & 0.85          & 0.86          \\ \hline
        ResNet-101          & 44 M            & 84.19           & 0.88          & 0.87          & 0.87          \\ \hline
        Alexnet             & 60 M           & 88.64           & 0.84          & 0.83          & 0.83          \\ \hline
        VGG-19              & 144 M           & 86.71           & 0.86          & 0.87          & 0.87          \\ \hline
    \end{tabular}%
    }
\end{table}

\begin{figure*}
\center

\includegraphics[width=0.3\textwidth]{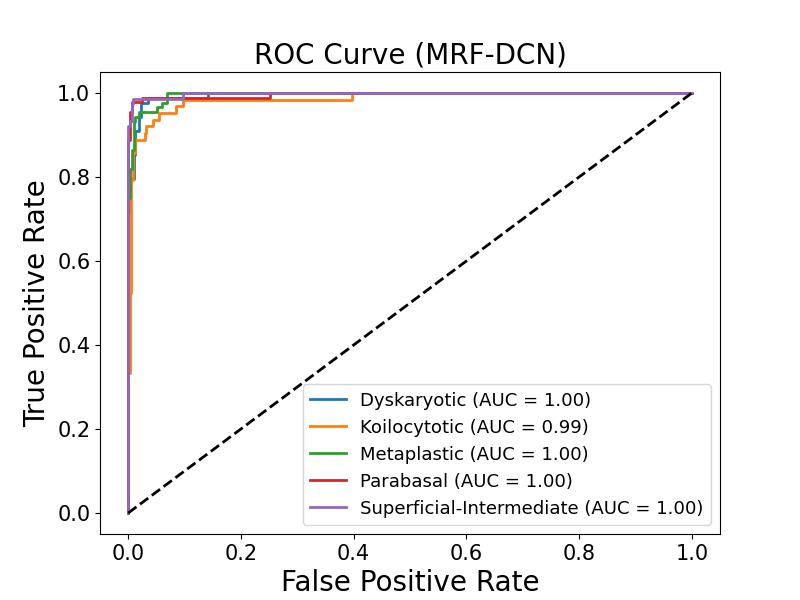}\hspace{0.5em}
\includegraphics[width=0.3\textwidth]{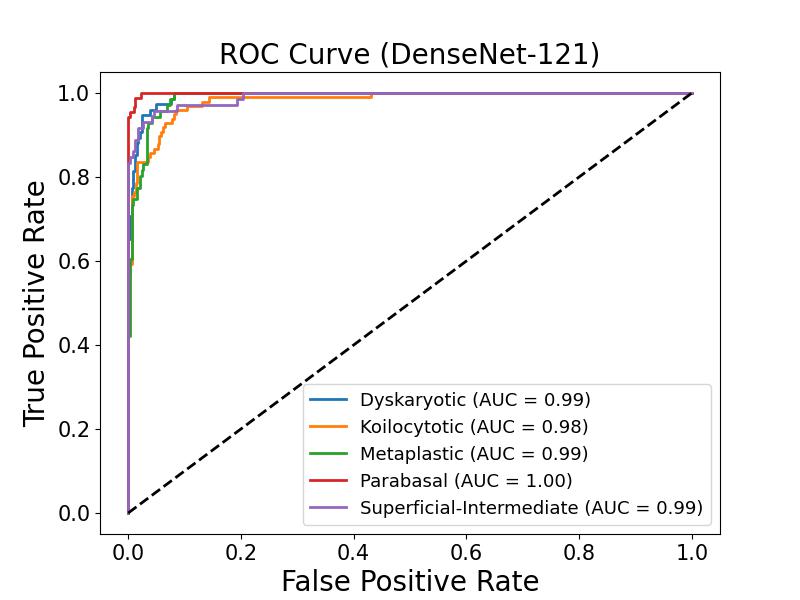}\hspace{0.5em}
\includegraphics[width=0.3\textwidth]{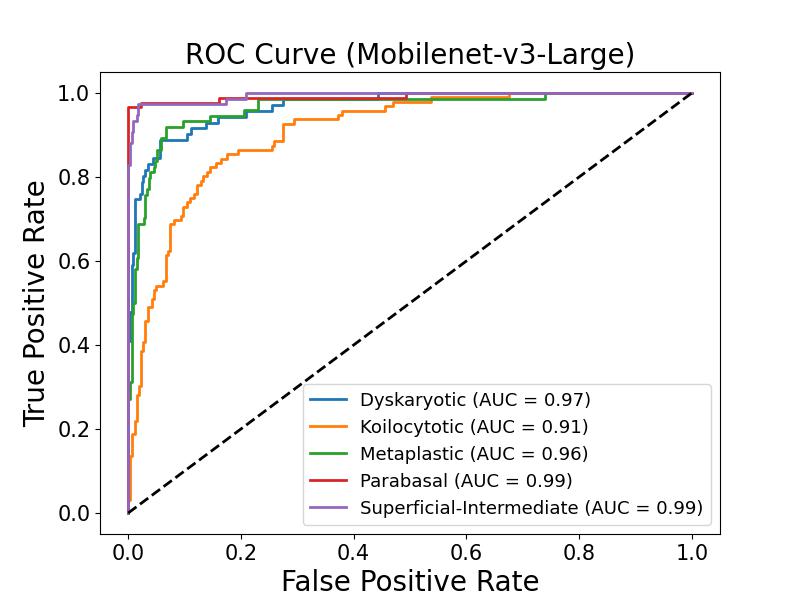}\\[0.5ex]

\includegraphics[width=0.3\textwidth]{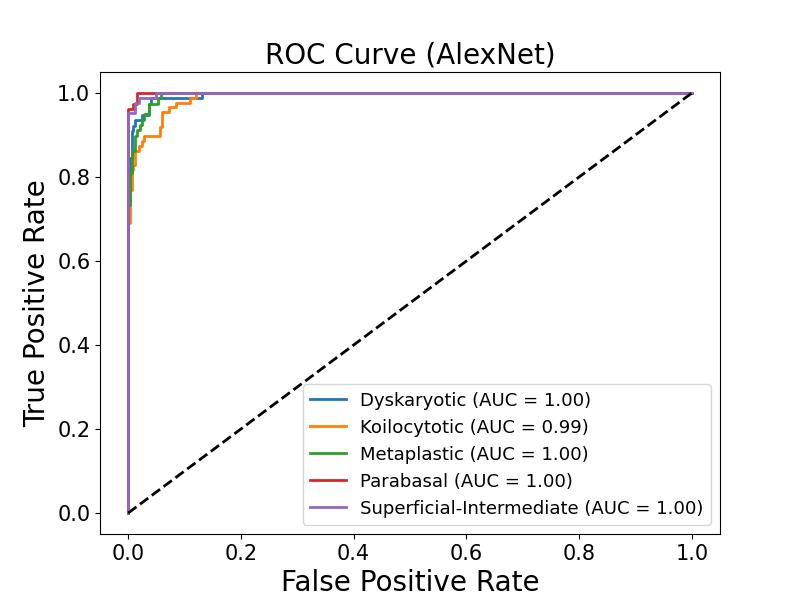}\hspace{0.5em}
\includegraphics[width=0.3\textwidth]{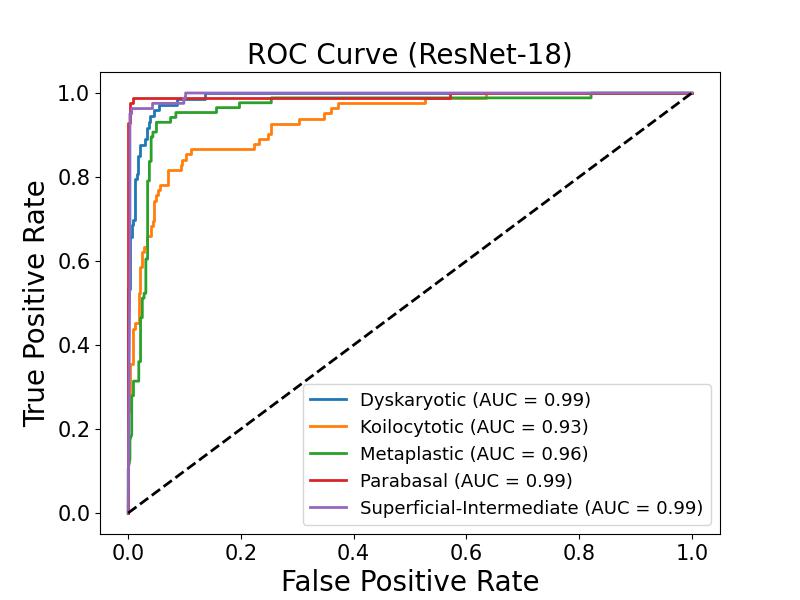}\hspace{0.5em}
\includegraphics[width=0.3\textwidth]{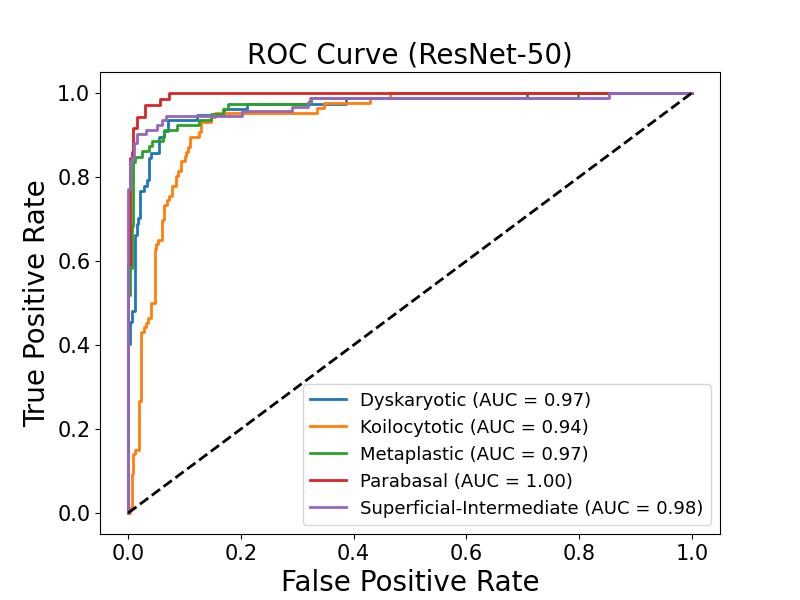}\\[0.5ex]

\includegraphics[width=0.3\textwidth]{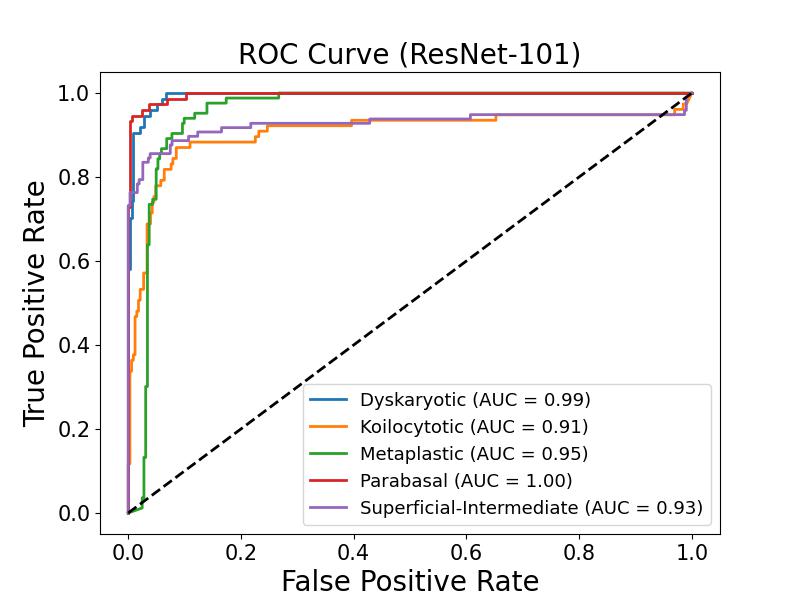}\hspace{0.5em}
\includegraphics[width=0.3\textwidth]{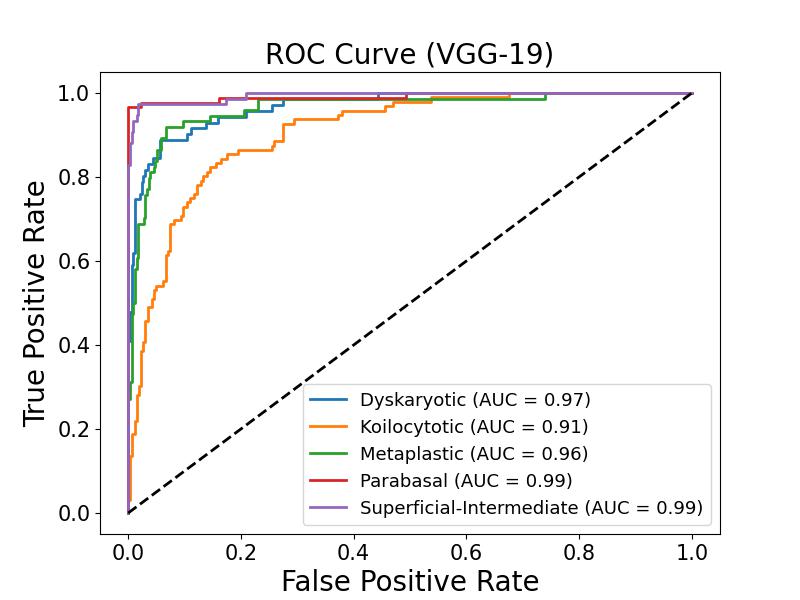}

\vspace{1ex}
\caption{ROC curves for various DL models evaluated on test set.}
\label{fig:roc_curves}
\end{figure*}

The proposed model's performance was meticulously evaluated and compared with existing models, achieving an accuracy (Acc) of 91.28\%, precision (Pre) of 0.91, recall (Rec) of 0.91 and F1-score(F1) of 0.91. The comparison, detailed in Table \ref{tab:comp_mrcnn} and Fig. \ref{fig:roc_curves}, shows that our model achieved results on par with the top-performing models in the field. The proposed MRF-DCN achieved outstanding performance among all evaluated models, with an AUC of 1.00 for four out of five classes and 0.99 for the remaining class. Its ROC curves closely align with the top-left corner, indicating near-perfect classification capability. MRF-DCN consistently demonstrates superior discrimination power across all cell classes compared to other state-of-the-art models. Our proposed model has 1.7 million learnable parameters, approximately 85 times fewer than the VGG-19 model \cite{vgg19}, 25 times fewer than the Resnet-101 model \cite{resnet}. This validation underscores the potential of our multi-scale approach to contribute meaningfully to advancing cell classification techniques. Moreover, our model's reduced number of parameters translates to faster training times and lower computational costs, making it an attractive option for practical applications.

\subsection{MTL}

This subsection explains the MTL technique adopted for cervical cancer diagnosis. As outlined earlier, the approach involves leveraging a modified UNet architecture, depicted in Fig.\ref{fig:mtl_unet}, specifically tailored to handle the dual objectives of segmentation and classification. The training process incorporates a composite loss function meticulously designed to optimize the network for both tasks simultaneously. For segmentation, we employ the binary cross-entropy loss function, defined in Eq. \ref{seg_bce_loss},
\begin{equation} \label{seg_bce_loss}
    \mathcal{L}_{\text{seg}} = - \frac{1}{N} \sum_{i=1}^{N} \left[ y_i \log (\hat{y}_i) + (1 - y_i) \log (1 - \hat{y}_i) \right],
\end{equation}
where $y_i$ is the ground truth label, $\hat{y}_i$ is the predicted probability, and $N$ is the total number of pixels. For classification, we use the categorical cross-entropy loss function, defined in Eq. \ref{cls_ce_loss},
\begin{equation} \label{cls_ce_loss}
    \mathcal{L}_{\text{cls}} = - \sum_{j=1}^{C} y_j \log (\hat{y}_j),
\end{equation}
where $C$ represents the number of classes. The final multi-task loss function is a weighted combination of both losses as defined in the Eq. \ref{comb_loss}
\begin{equation} \label{comb_loss}
    \mathcal{L}_{\text{total}} = \lambda_{\text{seg}} \mathcal{L}_{\text{seg}} + \lambda_{\text{cls}} \mathcal{L}_{\text{cls}},
\end{equation}
where $\lambda_{\text{seg}}$ $\in [0,1] $ and $\lambda_{\text{cls}}$ $\in [0,1] $ are hyperparameters that control the relative importance of each task. These loss components are strategically combined to guide the gradient-based optimization during backpropagation, ensuring the network’s parameters are iteratively refined to achieve a balanced performance across both tasks. This integration enhances the model’s capability to extract discriminative features while maintaining robust segmentation accuracy. The model achieved an IoU score of 0.83 for segmentation and a classification accuracy of 90\%, demonstrating its effectiveness in both tasks.

\vspace{-5mm}
\subsection{Risk Assessment}
This subsection delves into the intricacies of the RA process. As articulated previously, ML models were trained to rigorously evaluate whether the features extracted from cell images using the trained DL model exhibit sufficient discriminative capability. Our results demonstrated that the feature vectors were highly distinguishable, achieving more than 95\% accuracy across all ML models shown in Table \ref{tab:ra_fc}. Such outstanding performance underscores the efficacy of the DL model in generating a robust and highly discriminative feature set crucial for accurate classification and risk prediction.

\begin{table}[htp]
    \caption{Performance Measures of ML Models in Distinguishing Extracted Features for RA}
    \label{tab:ra_fc}
    \centering
    \resizebox{\columnwidth}{!}{%
    \begin{tabular}{|c|c|c|c|c|c|c|c|}
        \hline
        \textbf{Model} & \textbf{Acc} & \multicolumn{2}{c|}{\textbf{Precision}} & \multicolumn{2}{c|}{\textbf{Recall}} & \multicolumn{2}{c|}{\textbf{F1-Score}} \\ 
        \cline{3-8}
                       &                        & \textbf{M} & \textbf{W} & \textbf{M} & \textbf{W} & \textbf{M} & \textbf{W} \\ 
        \hline
        LR             & 95.20                 & 0.89       & 0.95       & 0.88       & 0.95       & 0.88       & 0.95       \\ 
        KNN            & 95.33                 & 0.89       & 0.95       & 0.88       & 0.95       & 0.88       & 0.95       \\ 
        Naive Bayes    & 95.16                 & 0.89       & 0.95       & 0.87       & 0.95       & 0.88       & 0.95       \\ 
        SVM            & 95.09                 & 0.89       & 0.95       & 0.87       & 0.95       & 0.88       & 0.95       \\ 
        SVM-Kernel     & 95.70                 & 0.89       & 0.95       & 0.87       & 0.95       & 0.88       & 0.95       \\ 
        RF             & 95.24                 & 0.89       & 0.95       & 0.88       & 0.95       & 0.88       & 0.95       \\ 
        DT             & 94.91                 & 0.88       & 0.95       & 0.87       & 0.95       & 0.87       & 0.95       \\ 
        \hline
    \end{tabular}%
    }
\vspace{-3mm}
\end{table}

These highly discriminative feature vectors are employed to evaluate the transition risk from a normal cell to an abnormal one. These extracted feature vectors are utilized in a probabilistic approach, as previously outlined, to estimate the likelihood of each class. This method computes probability scores for all classes based on the feature vectors. The class with the highest probability score is the most likely outcome, indicating a greater chance of progression into that particular class. Additionally, cosine similarity is computed between the test sample and the mean vector of each class. Generally ranging from -1 to 1, cosine similarity quantifies the closeness of feature vectors. Based on evaluations performed on the SIPaKMeD dataset, a cosine similarity greater than 0.65 indicates a high probability of progression into that class. Using the probabilistic approach previously delineated, in conjunction with the cosine similarity metric, the unseen samples can be seamlessly classified, and their similarity to all classes quantified.

This methodology was performed on a SIPaKMeD dataset, a retrospective dataset consisting of data collected and stored from past events. By exploiting this historical data, our method serves as a prognostic tool for cervical cancer detection, offering valuable insights into the potential progression of cellular abnormalities. These observations align with the broader objectives of RA, facilitating a prognosis framework that can guide clinicians in identifying high-risk cases and tailoring early intervention strategies to improve patient outcomes.

\vspace{-5mm}
\section{Conclusion} \label{conclusion}

In this research, we proposed a comprehensive framework addressing three critical tasks in cervical cell analysis: segmentation, classification, and RA. This article delivers two robust DL-oriented models and a new method for RA. For the classification task, we introduced a novel CNN architecture, MRF-DCN. This novel model demonstrated remarkable accuracy of 91.28 \% and efficiency in cell categorization, achieving state-of-the-art performance with learnable parameters 85 times less than the VGG-19 model's parameters and 25 times less than the Resnet-101 model's parameters. We proposed an MTL technique for both the tasks segmentation and classification, achieving an IoU score of 0.83 and 90\% accuracy, showcasing its capability to handle both segmentation and classification tasks simultaneously. Lastly, we developed an RA method to predict the progression of non-cancerous cells to malignant forms, which can be utilized for the prognosis of cervical cancer. This method evaluated the SIPaKMeD dataset as well as the Mendeley LBC dataset.

The limitation of our study, MRFDCN, lies in its fixed scaling mechanism. The model processes images at three predefined resolutions, restricting its adaptability to the original resolution of images—the model's fixed resolution processing limits general application to various medical imaging scenarios and test dataset settings. Despite this, our method's successful implementation and validation demonstrate the strength of DL models in medical image analysis and offer a promising foundation for future developments in early cancer detection and treatment. Future work will adapt MRF-DCN for MTL by implementing dynamic multi-scale strategies which can adjust to image resolution while extracting features from diverse spatial scales. Moreover, integrating dynamic weighting mechanisms to balance losses between tasks could improve performance and synergy, particularly in scenarios with task imbalance or conflicting objectives. We also plan to design lightweight segmentation models and compact architectures capable of efficiently performing both segmentation and classification while preserving high accuracy on high-resolution medical images.

\bibliographystyle{IEEEtran}

\bibliography{references}

\end{document}